\documentclass[pra,showpacs,showkeys]{revtex4}
\usepackage{amsfonts,amsmath,color}
\usepackage[dvips]{graphics}
\begin{document}

\newcommand{\ket}[1]{\mathop{\left|#1\right>}\nolimits}       % |ket>
\newcommand{\bra}[1]{\mathop{\left<#1\,\right|}\nolimits}       % <bra|
\newcommand{\Tr}[1]{\mathop{{\mathrm{Tr}}_{#1}}}              % Trace
\newcommand{\braket}[2]{\langle #1 | #2 \rangle}
\newcommand{\ketbra}[2]{| #1\rangle\!\langle #2 |}
\newcommand{\nn}{\nonumber}
\newcommand{\vsp}{\vspace{.5cm}}

\title{Entanglement enhancement and postselection for two atoms interacting with thermal light}

\author{K. Br\'adler}
\author{R. J\'auregui}
\affiliation{Instituto de F\'{\i}sica, Apdo. Postal 20-364,
M\'exico 01000, M\'exico}

\email{kbradler@epot.cz, rocio@fisica.unam.mx}

\date{\today}

\begin{abstract}
The evolution of entanglement for two identical two-level atoms
coupled to a resonant thermal field is studied for two different
families of input states. Entanglement enhancement is predicted
for a well defined region of the parameter space of one of these
families. The most intriguing result is the possibility of
probabilistic production of maximally entangled atomic states even
if the input atomic state is factorized and the corresponding
output state is separable.
\end{abstract}



\maketitle

Entanglement of atomic systems is a promising resource for
performing various computational and communication tasks
originated in quantum information theory. Recent experimental
progress has been achieved in the field of creation of
multipartite highly nonclassical states in a linear Paul
trap~\cite{ion1_nature,ion2_nature}. Production of such states
requires a nontrivial implementation of a sequence of nonlocal
operations (unitary gates) that entangle originally uncorrelated
atoms in a desired way. Apart from these sophisticated methods for
manipulating the atomic states with the help of a precisely tuned
laser field, it has been predicted that originally independent
atoms may become entangled after the interaction with a thermal
electromagnetic field. At first sight this may seem
counterintuitive because thermal light is considered to be a
decoherence source. But indeed, if we stay in the quite modest
case of entangling just two atoms, Kim~{\it
et~al.}~\cite{thermal_light} as well as Bose~{\it
et~al.}~\cite{thermal_light_ent} and An~{\it
et~al.}~\cite{thermal_light_ent1} have reported the occurrence of
entanglement of initially separable mixed states after the
interaction with thermal light. These studies belong to a broader
group of works  on the behavior of an atomic ensemble coupled to
different kinds of bosonic environment~\cite{fock,whitenoise}.

Loosely speaking, all previous studies of the coupling of an
atomic system to a thermal
bath~\cite{thermal_light,thermal_light_ent,thermal_light_ent1,thermal_others}
concluded with the interesting result that entanglement not only
does not disappear after the interaction but it can be even
created from originally separable states. In this work we go
further. In the framework of the model describing the interaction
of an atomic ensemble and a single mode photon field worked out by
Tavis {\it et~al.}~\cite{tavis} we study two physically relevant
families of initial pure atomic states coupled to a thermal bath.
The states can be tuned by changing its Schmidt number
(characterizing the degree of entanglement of initial states) from
factorized states to all four maximally entangled states (Bell
states). The calculations give  quite unexpected results that, to
our knowledge, have not been reported so far. It is found that for
an input state of the form
$\ket{\Psi_\beta}=\sin\beta\ket{00}+\cos\beta\ket{11}$ and for a
sufficiently 'hot' environment, there is a stable nonzero
probability of producing a maximally entangled state irrespective
of the value of $\beta$ and, what is more interesting, despite of
the fact that the resulting state after the interaction is
separable. This procedure is usually called postselection. Thus we
present here a probabilistic source of maximally entangled atomic
pairs.

A detailed study of the  family of input atomic states
$\ket{\Phi_\beta}=\sin\beta\ket{01}+\cos\beta\ket{10}$ shows that
entanglement of output states is higher in comparison with input
states for a wide interval of $\beta$ values. The picture started
in Refs.~\cite{thermal_light,fock} is now completed by showing
that the entangling properties of a thermal field prevail even at
very low temperatures, so that vacuum fluctuations can generate
partially entangled states from initially factorized states.

The structure of the paper is as follows. In
Section~\ref{measures} we introduce the entanglement of formation
(EOF)~\cite{EOF}. This entanglement measure is calculated for
general two-qubit states with the help of the analytical
expression for the concurrence~\cite{wootters}. In
Section~\ref{main} we describe the Tavis-Jaynes-Cummings model
setting the stage for the main results presented in two
subsections where the evolution of thermal light with two
different atomic input configurations $\ket{\Phi_\beta}$ and
$\ket{\Psi_\beta}$ is studied. The details about the calculations
are given in Appendix.

\section{Entanglement measures}
\label{measures}

Most axioms required for entanglement measures~\cite{measures} are
accomplished by the entanglement of formation~\cite{EOF} (EOF)
defined by
\begin{equation}\label{EOF}
    E_F({\varrho})=\inf_{\varrho=\sum_ip_i\ketbra{\varphi}{\varphi}_i}
    \sum_ip_iE\left(\ketbra{\varphi}{\varphi}_i\right),
\end{equation}
where
$E\left(\bullet\right)=S(\Tr{1(2)}\left[\ketbra{\varphi}{\varphi}_i\right])$
is called the entropy of entanglement (the von Neumann entropy of
the state of interest that is traced over one of its subsystems).
The EOF is generally uneasy to calculate even for the lowest
dimensional systems. However, Hill and Wootters
proved~\cite{wootters} that the EOF can be calculated analytically
in the case of an arbitrary two qubit state $\varrho$. They showed
that
\begin{equation}\label{Shannon}
    E_F({\varrho})=h\left(\frac{1}{2}\left(1+\sqrt{1-C^2(\varrho)}\right)\right),
\end{equation}
where $h(x)=-x\log_2(x)-(1-x)\log_2(1-x)$ is the Shannon entropy
and
\begin{equation}\label{con}
    C(\varrho)=\max\{0,\lambda_1-\lambda_2-\lambda_3-\lambda_4\}.
\end{equation}
The parameters $\lambda_i$ in Eq.~(\ref{con}) are eigenvalues in
descending order of the square root of the matrix
\begin{equation}\label{R}
    \tilde R(\varrho)=\varrho\left(\sigma_y\otimes\sigma_y\right)
    \varrho^*\left(\sigma_y\otimes\sigma_y\right)
\end{equation}
with $*$ representing the complex conjugation in the standard
basis and $\sigma_y$ is the $y$-Pauli matrix. $C$ is called the
concurrence and since the EOF is monotonous in $C$, the
concurrence can also be considered as the entanglement measure.
Nevertheless, as emphasized in~\cite{plenio}, the proper
entanglement measure is the EOF. Using the concurrence for this
purpose could cause confusion when investigating entanglement of
systems with dimensionality $n\times n$ for $n\ne 2$ since in that
case the definition of concurrence is not unique.

Also in some cases, e.g. in~\cite{thermal_light}, the
negativity~\cite{negativity} as an entanglement measure is
preferred. The reason is that it is easier to calculate (tracing
over a partially transposed density matrix). On the other hand,
the negativity does not coincide with the entropy of entanglement
introduced above for pure bipartite states where it is the unique
measure of entanglement~\cite{uniqueness}. Rather
recently~\cite{CREN}, connection of the negativity with the
concurrence (and thus to the EOF that coincides with the entropy
of entanglement for pure bipartite states) was discovered in terms
of convex roof construction. Anyhow, in this paper the particular
forms of the density matrices allow us to calculate the EOF
directly from the concurrence.

The second note concerns the term Schmidt number. It characterizes
the degree of entanglement of input states but does not uniquely
determine the behavior after the interaction. In other words, even
if two input states have the same Schmidt number, for example two
Bell states, their evolution and thus entanglement after the
interaction may be completely different.

\section{Interaction of two atoms with thermal light}
\label{main}

The Hamiltonian  of two identical two-level atoms interacting with
a single-mode electromagnetic field in the dipole approximation
and standard notation is given by
\begin{eqnarray}\label{Hamiltonian}
    H&=&H_0+H_{int}\nonumber\\
    &=&\hbar\Omega\left(b^\dagger b+\frac{1}{2}\right)
    +\frac{\hbar\Omega}{2}\left(\sigma_z^{(1)}+\sigma_z^{(2)}\right)
    +\hbar g\left(\sum_{i=1}^2\sigma^{(i)}_+b+\sigma^{(i)}_-b^\dagger\right),
\end{eqnarray}
where resonance of the photon energy and the atomic level
splitting is assumed, and the rotating-wave approximation is used.
The analogous Hamiltonian for $N$-atoms can be analytically
diagonalized as already shown in Ref.~\cite{tavis}. Here we use
the dressed states formalism to obtain the eigenvalues and
eigenvectors. Since $\left[H_0,H_{int}\right]=0$, $H_{int}$
induces transitions only between the degenerate states of $H_0$
that, for a given field excitation number $n$, constitutes the
tetrad $\{\ket{n}\ket{11},\ket{n+1}\ket{10},\ket{n+1}\ket{01},
\ket{n+2}\ket{00}\}_{n}$. The bare states form a 'semilogical'
basis where, e.g., $\ket{n+1}\ket{10}$ represents the state with
$(n+1)$ photons, the first atom  in the excited state (logical
state one) and the second one in the ground state (logical state
zero). Expressing the interaction Hamiltonian in this basis we get
a block diagonal matrix with the $n$-th block given by
\begin{equation}\label{Hint}
H_{int}^{(n)}=\hbar g
\begin{pmatrix}
  0          & \sqrt{n+1} & \sqrt{n+1} & 0 \\
  \sqrt{n+1} & 0          & 0          & \sqrt{n+2} \\
  \sqrt{n+1} & 0          & 0          & \sqrt{n+2} \\
  0          & \sqrt{n+2} & \sqrt{n+2} & 0 \\
\end{pmatrix}.
\end{equation}
After normalization of the eigenvectors of this matrix, the
dressed basis states read
\begin{subequations}
\label{dressed_basis}
\begin{eqnarray}
\ket{1}_n &=&\sqrt{\frac{n+1}{2n+3}}\left[-\sqrt{\frac{n+2}{
n+1}}\ket{n}\ket{11}
                                   +\ket{n+2}\ket{00}\right]\\
\ket{2}_n &=& \sqrt{\frac{1}{2}}\left[\ket{n+1}\ket{10}-\ket{n+1}\ket{01}\right]\\
\ket{3}_n
&=&\sqrt{\frac{2}{4n+6}}\left[\sqrt{\frac{n+2}{n+1}}\ket{n}\ket{11}
                                   -\sqrt{\frac{2n+3}{2n+4}}\ket{n+1}\ket{10}
                                   -\sqrt{\frac{2n+3}{2n+4}}\ket{n+1}\ket{01}
                                   +\ket{n+2}\ket{00}\right]\\
\ket{4}_n &=&
\sqrt{\frac{n+2}{4n+6}}\left[\sqrt{\frac{n+2}{n+1}}\ket{n}\ket{11}
                                   +\sqrt{\frac{2n+3}{2n+4}}\ket{n+1}\ket{10}
                                   +\sqrt{\frac{2n+3}{2n+4}}\ket{n+1}\ket{01}
                                   +\ket{n+2}\ket{00}\right].
\end{eqnarray}
\end{subequations} In particular, the atomic state
$1/\sqrt{2}(\ket{10}-\ket{01})$ is a dark state.

\subsection{Initial Atomic State $\ket{\Phi_\beta}$}

\begin{figure}
\resizebox{18cm}{12cm}{\includegraphics{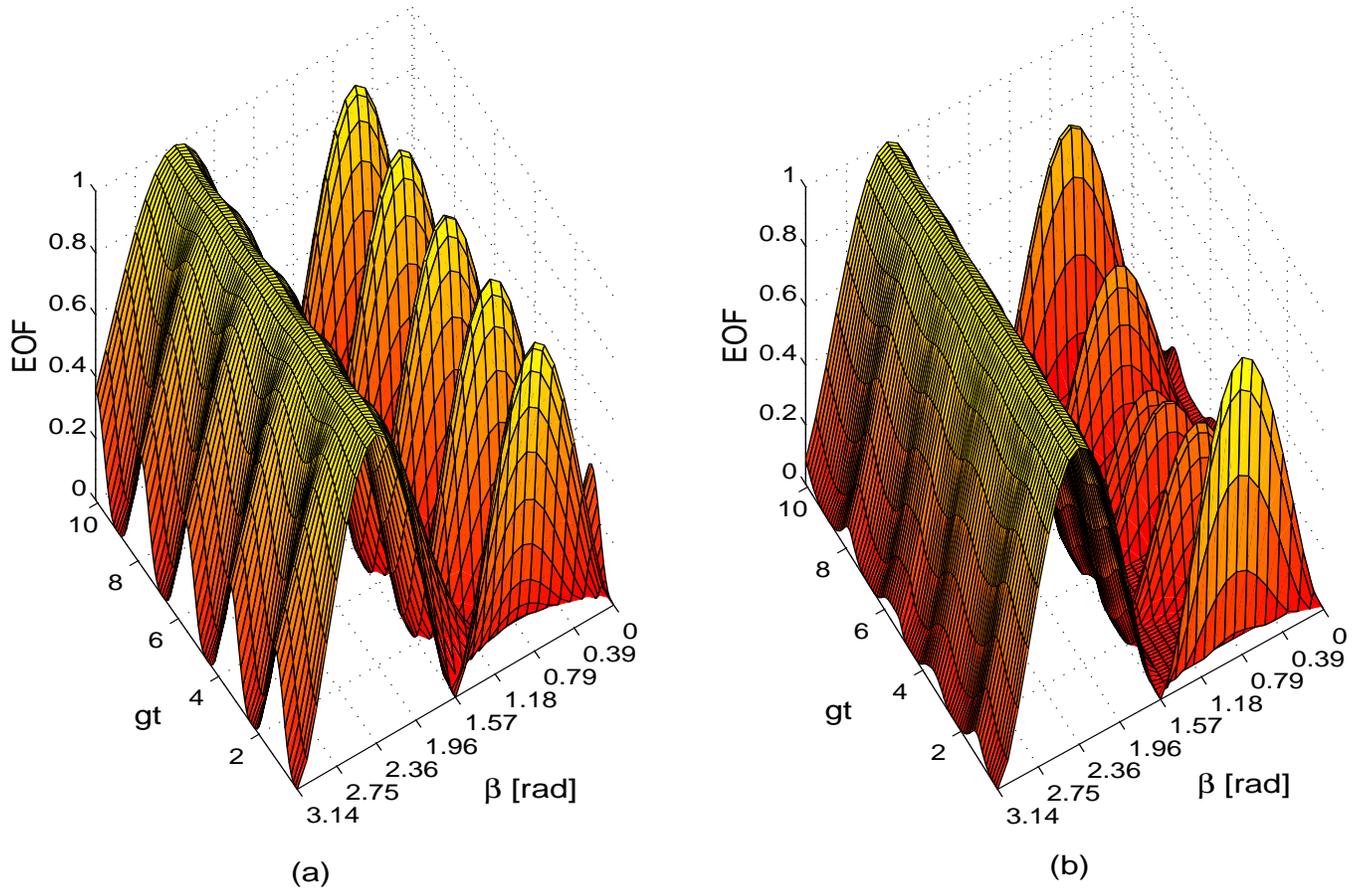}}
\caption{\label{eofb_phi_n0}The EOF of the output state given by
Eq.~(\ref{rho_Phi}) for interacting atomic states
$\ket{\Phi_\beta}$ with: (a) a zero photon field, and (b) a
thermal field with a nonzero average photon number ($\bar
n\approx0.64$) . A completely different behavior for
$0\leq\beta\leq\pi/2$ and $\pi/2\leq\beta\leq\pi$ can be seen. EOF
revivals are observed in the first interval where, generally,
entanglement of input states is devalued except for initially
factorized or weakly entangled states. On the other hand, in the
second interval the EOF is never lower  than its initial value. We
can observe significant values of the EOF for initially factorized
atomic states and $\bar n =0$.}
\end{figure}

In this section, the system light-atoms is assumed to be initially
in a factorized state
$\varrho^{(\Phi)}(\beta,0)=\varrho_{light}\otimes\varrho^{(\Phi)}_{atoms}$
where the atomic state is a pure state
\begin{equation}\label{Phi}
    \ket{\Phi_\beta}=\sin\beta\ket{01}+ \cos\beta\ket{10}
\end{equation}
with the Schmidt number $\sin\beta$ and the light considered to be
in a thermal state
\begin{eqnarray}\label{light}
    \varrho_{light}
    &=&\left[1-e^{-\hbar\Omega/kT}\right]e^{\hbar\Omega/2kT}\sum_{n=0}^\infty
    e^{-(n+1/2)\hbar\Omega/kT}\ketbra{n}{n}\nonumber\\
    &=&\sum_{n=0}^\infty \frac{e^{\bar n}\bar
    n^n}{n!}\ketbra{n}{n},
\end{eqnarray}
with $\bar n=\left[e^{\hbar\Omega/kT}-1\right]^{-1}$ the average
number of photons. After a direct but otherwise tedious
calculation the time evolved density matrix
$\varrho^{(\Phi)}(\beta,t)$ can be obtained. Since we are
interested in the entanglement behavior of the atom-atom system,
we trace over the electromagnetic field states. The resulting
time-dependent reduced density matrix is of the form
\begin{equation}\label{rho_Phi}
    \tilde\varrho^{(\Phi)}(\beta,t)
    =\begin{pmatrix}
      a_1 & 0 & 0 & 0 \\
      0 & a_2 &  a_3 & 0 \\
      0 & a_3 & a_4 & 0 \\
      0 & 0 & 0 & a_5 \\
    \end{pmatrix}.
\end{equation}
The specific expressions of the matrix elements can be found in
Appendix.

The structure of $\tilde\varrho^{(\Phi)}$ is very friendly for the
calculation of the concurrence with Eq.~(\ref{con}) and
consequently the EOF. The result is illustrated in
Fig.~\ref{eofb_phi_n0} where  oscillations of the EOF are
noticeable for all $\beta$. There is a $\beta_0\ll1$ that depends
on $ \bar n$ such that for the subinterval
$\beta_0<\beta<\pi/2-\beta_0$ the minimum value of the EOF is zero
and the local maxima (revivals) are always smaller than the
initial value for $\bar n>0$ and equal for $\bar n=0$. Thus, the
EOF of the atomic entangled states is reduced in the presence of
thermal light. As the temperature increases the temporal width of
the revivals decreases. A brief inspection of the
 picture reveals other interesting aspects of interacting light
 with this family of states $\ket{\Phi_\beta}$ in
 the second interval of $\beta$, $\pi/2 <\beta<\pi$ . Due to the fact that
 $1/\sqrt{2}(\ket{10}-\ket{01})$ is a dark state the behavior  is completely different. The EOF
 exhibits very soft oscillations and is never smaller than its
 initial value. We thus see that in this region, the resulting
 state is stable under the interaction with thermal light. This
 could be an important hint for quantum engineers (and protocol
 designers generally)  telling them that there exists an important
 class of initial atomic states where the interaction with a
 thermal field means not only no loss of entanglement but even its
 enhancement without strong fluctuations. The works presented so
 far on this topic (e.g.
 Refs.~\cite{thermal_light,thermal_light_ent}) were limited only to
 input atomic states with small values of the Schmidt number
 $0<\beta<\beta_0\ll1$ and recognized the presence of the dark
 state but do not discuss the stability of entanglement in the
 whole range of $\beta$.

For $\bar n=0$ the matrix elements of Eq.~(\ref{rho_Phi}) have a
particulary simple form
\begin{subequations}\label{rho_Phi_n0}
\begin{eqnarray}
a_1 &=& \frac{1}{4}(1+\sin2\beta)(1-\cos2\alpha_{-1}t)\\
a_2 &=& \frac{1}{8}(1+\sin2\beta)(1+\cos2\alpha_{-1}t)
        -\cos2\beta\cos\alpha_{-1}t+\frac{1}{4}(1-\sin2\beta)\\
a_3 &=& \frac{1}{8}(1+\sin2\beta)(1+\cos2\alpha_{-1}t)-\frac{1}{4}(1-\sin2\beta)\\
a_4 &=& \frac{1}{8}(1+\sin2\beta)(1+\cos2\alpha_{-1}t)
        +\cos2\beta\cos\alpha_{-1}t+\frac{1}{4}(1-\sin2\beta)\\
a_5 &=& 0.
\end{eqnarray}
\end{subequations}
 By examining the density matrix for $\beta=0$ we may exactly see
the influence of the vacuum fluctuations on a factorized input
state. We see that there is a $50\%$ probability of finding the
system with both atoms in their ground state and a single photon,
while there is a $50\%$ probability that this photon has been
reabsorbed by one of the atoms leaving the atomic system in the
entangled state. As a consequence, spontaneous decay may be
regarded as a causative mechanism for inducing entanglement
between the two two-level atoms when their initial state
$\ket{\Phi_\beta}$ is practically unentangled ($\beta\simeq 0$) or
to enhance it for different $\beta$ in the interval
$\pi/2\leq\beta\leq\pi$.

\begin{figure}[t]
\resizebox{15cm}{10cm}{\includegraphics{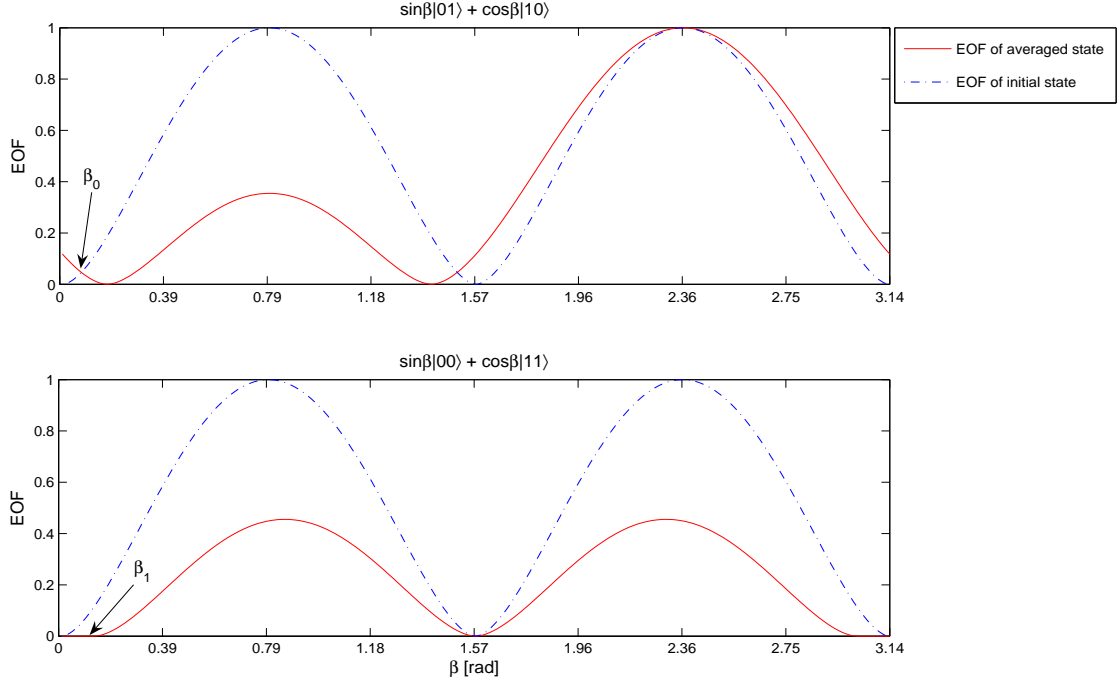}}
\caption{\label{fig_aver} The continuous lines in both plots are
the EOF for mixtures of the output states occurring within the
time interval where the EOF is periodic ($\bar n=0$).  The
dash-dotted lines correspond to the EOF for $t=0$ (initial
states). In the upper plot the interval
$\beta_0<\beta<\pi/2-\beta_0$ with $\beta_0=1/2\arcsin({1/7})$
demarcates the region with lower entanglement in comparison to
corresponding input states. In the bottom plot we can see that
entanglement is not present in the interval $(-\beta_1,\beta_1)$,
with $\beta_1=\arctan(1/8)$, hence disappearing even for initially
weakly entangled states.}
\end{figure}

The fact that EOF exhibits strong oscillations as already
illustrated may limit even its evaluation. Evaluation of the
entanglement requires knowledge of the matrix elements and these
cannot be determined just from one copy of an unknown quantum
state. One possibility is to have many copies of the same unknown
state and perform quantum state tomography. Recent findings show
that the tomography is not
necessary~\cite{ent_exp_measure1,ent_exp_measure2} but still does
hold that it is not possible to measure the entanglement from just
one copy of a state. Besides, if the average number of photons is
so large that the time window for picking the desired state is too
long  compared to the length of the oscillations, the measurement
could be  interpreted as a random pick from the set of all
possible outputs and a proper time average must be performed.

 Let us study this question for the electromagnetic field initially
  in the vacuum
state as an illustration. In this simple case, the natural period
is determined by the commeasurable Rabi frequencies $\sqrt{8}g$
and $\sqrt{2}g$
\begin{equation}\label{average_over_gt}
    \tilde\varrho^{(\Phi)}(\beta)
    =\frac{1}{\sqrt{2}\pi}\int_0^{\sqrt{2}\pi}\varrho(\beta,t){\rm
     d}(gt)
    =\frac{1}{8}\begin{pmatrix}
      2(1+\sin2\beta) & 0 & 0 & 0 \\
      0 & 3-\sin2\beta & 3\sin2\beta-1 & 0 \\
      0 & 3\sin2\beta-1 & 3-\sin2\beta & 0 \\
      0 & 0 & 0 & 0 \\
    \end{pmatrix}.
\end{equation}
The resulting EOF can be seen in the upper plot of
Fig.~\ref{fig_aver} and, we are able to confirm the qualitative
conclusions about the entanglement behavior based on
Fig.~\ref{eofb_phi_n0} for the averaged EOF. As mentioned above,
we can observe nonzero entanglement for initially factorized
states. For entangled states with $\beta_0<\beta<\pi/2-\beta_0$
the averaged EOF is lower than for the input states. In general,
for the remaining $\beta$ the EOF average is higher than the
initial value of the EOF.

The extension of this averaging process for the case of an
interaction with thermal light with $\bar n>0$ requires the
identification of a natural time scale since in this case an
infinite number of the incommensurable Rabi frequencies
$\alpha_{n-1}=2g\sqrt{n+1/2}$ determines the evolution of the
system. One such scale could be found by the following procedure.
Writing the  summations in the expression of
$\tilde\varrho^{(\Phi)}(\beta,t)$ in  an integral form using the
Abel-Plana formula, it can be shown that the adimensional
parameter $gt\sqrt{2kT/\hbar\Omega}$ is especially important to
understand the evolution of the system. In fact, it can be used to
define a natural time $\tau_0=g^{-1}\sqrt{\hbar\Omega/2kT}$ to
distinguish short and long time effects of the interaction (see
Appendix for more details). If the average procedure from above is
applied, $\tau_0$ could also be used for this purpose.

\subsection{Initial Atomic State $\ket{\Psi_\beta}$}

Following the same reasoning as in the previous subsection, but
for an initial state of the form
\begin{equation}\label{Psi}
    \ket{\Psi_\beta}=\sin\beta\ket{00}+\cos\beta\ket{11}
\end{equation}
interacting with light in a thermal state~(\ref{light}), we arrive
to a reduced density matrix with the structure
\begin{equation}\label{rho_Psi}
    \tilde\varrho^{(\Psi)}(\beta,t)
    =\begin{pmatrix}
      b_1 & 0 & 0 & b_2 \\
      0 & b_3 & b_3 & 0 \\
      0 & b_3 & b_3 & 0 \\
      b^*_2 & 0 & 0 & b_4 \\
    \end{pmatrix}.
\end{equation}
The particular expressions for the matrix elements are given in
Appendix. As illustrated in Fig.~\ref{eofb_psi_n0}, in this case a
resonant thermal electromagnetic field is not able to induce, in
general, atomic entanglement for input factorized atomic states,
nor to enhance it for initial already entangled states. An
exception is provided for the states with $\beta \simeq \pi/2$.
There, a very slight EOF enhancement is observed for  $0<\bar
n\lesssim 2$ . This effect is discussed at length in
Ref.~\cite{thermal_light} and present, but barely observable, in
Fig.~\ref{eofb_psi_n0}. For the other values of $\beta$, at most,
partially entangled input states exhibit revivals of the EOF.
\begin{figure}
\resizebox{18cm}{12cm}{\includegraphics{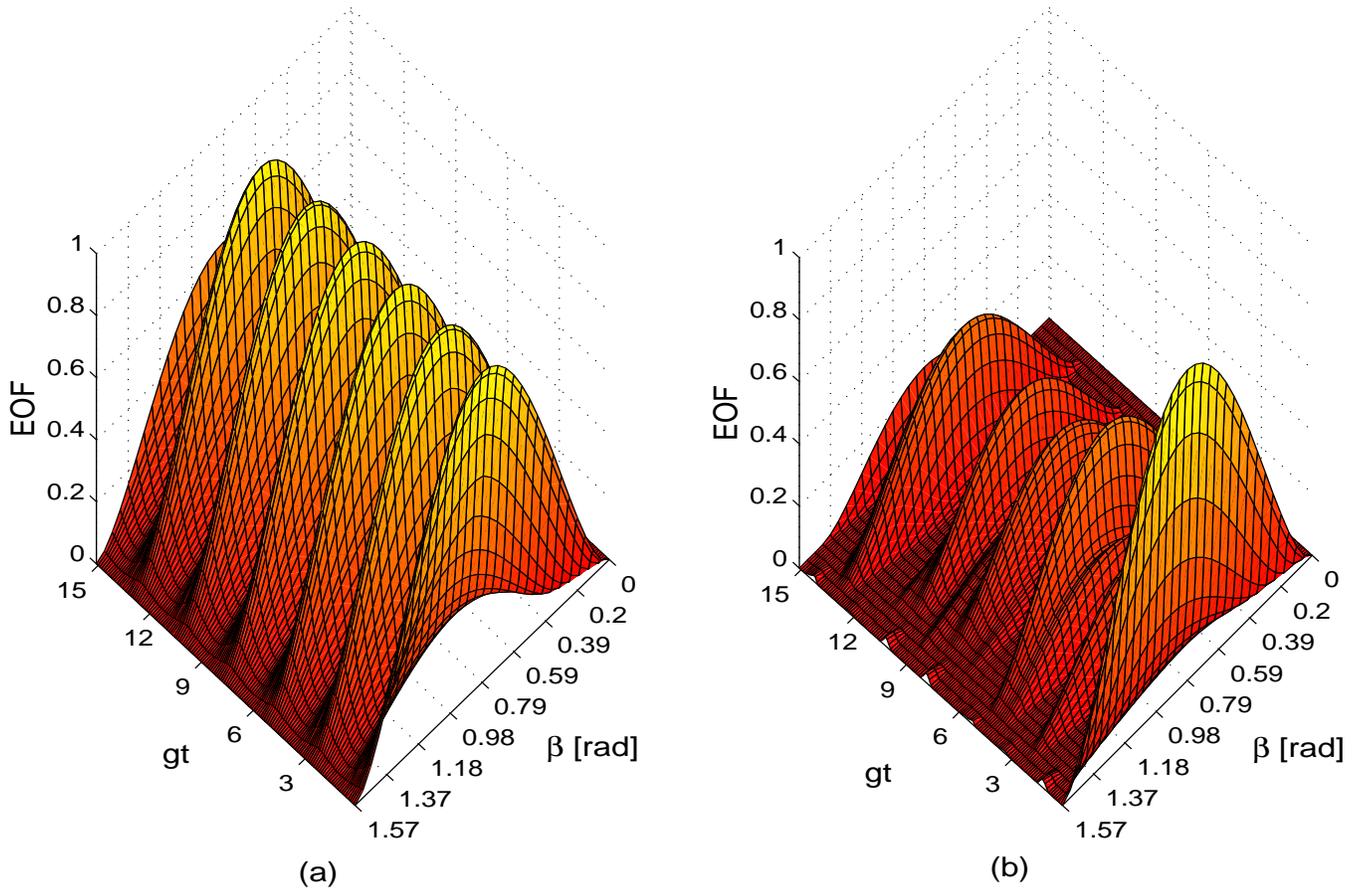}}
\caption{\label{eofb_psi_n0} The EOF of output
state~(\ref{rho_Psi}) for interacting atomic states
$\ket{\Psi_\beta}$ with (a) a zero photon field, (b) a thermal
field with $\bar n\approx0.64$. Notice the small enhancement of
the EOF for $\beta \sim \pi/2$ where $\bar n$ was chosen to
provide a maximum value of the enhancement.}
\end{figure}
Thus, in comparison to the input state $\ket{\Phi_\beta}$, the
entanglement properties of a thermal field are not so striking.
This qualitative observation is confirmed for $\bar n=0$ by the
mixture calculation in the spirit of the previous section. It
gives us
\begin{equation}\label{average_over_gt1}
    \tilde\varrho^{(\Psi)}(\beta)
    =\frac{1}{12}\begin{pmatrix}
      4(2-\cos2\beta) & 0 & 0 & 4\sin2\beta \\
      0 & \cos^2\beta & \cos^2\beta & 0 \\
      0 & \cos^2\beta & \cos^2\beta & 0 \\
      4\sin2\beta & 0 & 0 & 6\cos^2\beta \\
    \end{pmatrix}.
\end{equation}
The resulting EOF is depicted in Fig.~\ref{fig_aver} (bottom
plot). We see that in the mixture sense, there is not only a
separable output for initially unentangled states, as discussed
above, but also for some of the input partially entangled states.
Also, if we compare the EOF for the initial state (dash dotted
line) and evolved state (continuous line) the output states are
much less entangled in general.

\begin{figure}[t]
\resizebox{15cm}{10cm}{\includegraphics{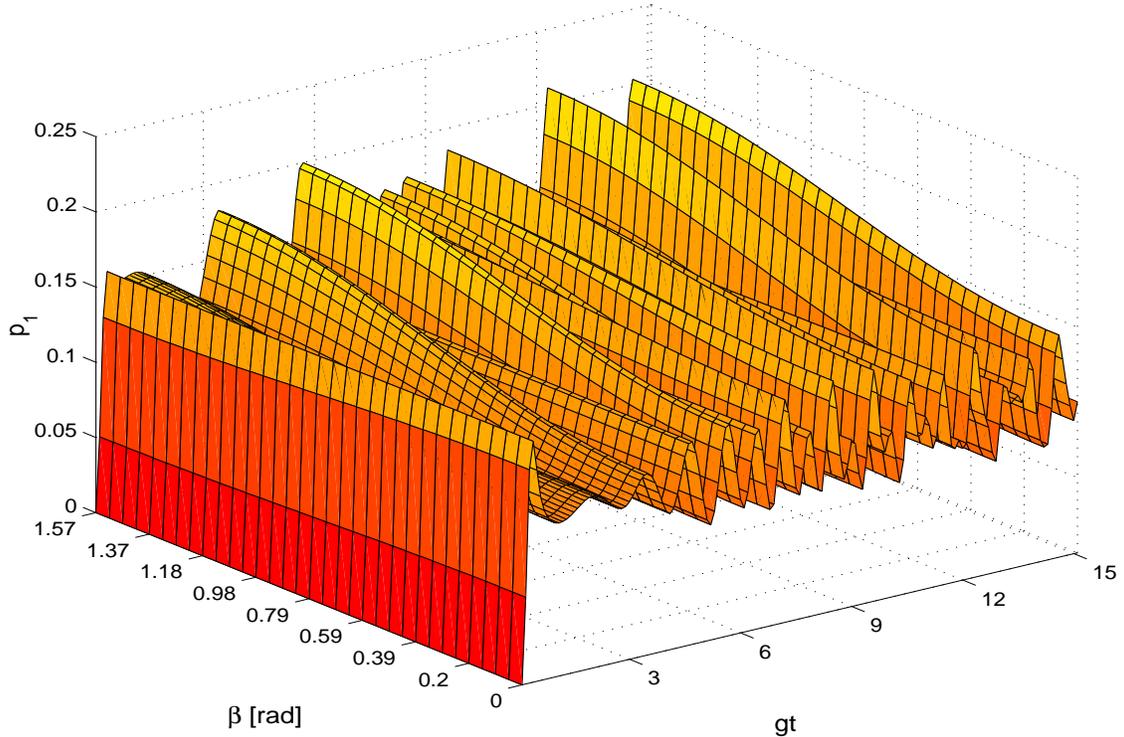}}
\caption{\label{fig_Bcoeff1} Depicted values of parameter $p_1$
for $\bar n\approx5.9$ which determines the probability of the
successful postselection of a maximally entangled state
$1/\sqrt{2}(\ket{10}+\ket{01})$ from outgoing
state~(\ref{rho_Psi}). It is possible to postselect the state
irrespective of $\beta$ and thus it can be done even if the output
state is separable.}
\end{figure}

Nevertheless, returning to the evolved state~(\ref{rho_Psi}),
another interesting aspect appears. Eq.~(\ref{rho_Psi}) can be
written as
\begin{equation}\label{Psi_out}
\tilde\varrho^{(\Psi)}(\beta,t)=p_1\varrho_1+(1-p_1)\varrho_2
\end{equation}
where $p_1\equiv b_3$ and
\begin{eqnarray}\label{Psi_decomp}
\varrho_1&=&\frac{1}{2}\ketbra{01+10}{01+10}\nonumber\\
\varrho_2&=&\frac{1}{b_1+b_4}(b_1\ketbra{00}{00}+b_4\ketbra{11}{11}
+b_2\ketbra{00}{11}+b_2^*\ketbra{11}{00})
\end{eqnarray}
and both states can be unambiguously discriminated. To show this,
let us consider a projectively measuring apparatus with three
 possible outputs. Two of them project into orthogonal subspaces
 spanned by $\ketbra{00}{00}$ and $\ketbra{11}{11}$ and the third
 one is the complement. Then, from the form of~(\ref{Psi_out})
 it follows that if we do not detect anything on the first and the
 second output the state is in the required maximally entangled
 state $\frac{1}{\sqrt{2}}\ket{01+10}$. The probability of this
 procedure is given by the parameter $p_1$. Thus, no additional
 quantum resource for the discrimination is needed
\footnote{If, from some reasons, we need to preserve both output
states $\varrho_1,\varrho_2$ intact we introduce an auxiliary
qubit $\ket{0}_{aux}$ and the unitary matrix
$T=CNOT_{13}CNOT_{23}$ where in $CNOT_{mn}$ $m$ and $n$ is a
control and a target qubit, respectively. This is useful for
non-demolition discrimination between $\varrho_1$ and $\varrho_2$
since under this transformation we get
$\tilde\varrho\otimes\ketbra{0}{0}_{aux}
    \stackrel{T}{\to}p_1\varrho_1\otimes\ketbra{1}{1}_{aux}
    +(1-p_1)\varrho_2\otimes\ketbra{0}{0}_{aux}$.
So that, by a projective measurement on the ancilla, $\varrho_1$
and $\varrho_2$ can be unambigously discriminated without
destroying them.}.

Therefore, even if the EOF is zero (i.e. when an output state is
separable) the entangled pair can be extracted in terms of
postselection. Also, it is interesting that the maximum value of
$p_1$ is almost the same not only for different input states (see
Fig.~\ref{fig_Bcoeff1}) but also for different average photon
numbers of thermal light. It is worth remarking that the higher
photon average number is used the less fluctuating the parameter
$p_1$ we get. This constant character of $p_1$ is broken only for
the case of a zero photon field where, not surprisingly, for an
input state $\ket{00}$ we cannot extract a maximally entangled
state by any means, see Eq.~(\ref{average_over_gt1}).

Note that a similar procedure cannot be applied to the resulting
state in Eq.~(\ref{rho_Phi}) from the previous subsection. There,
we are not able to find a decomposition of a density matrix in the
form $\varrho=\varrho_1\oplus\varrho_2$ as in the current case,
which ensures that if two arbitrary density matrices are positive
the third must be so.

\section{Conclusions}

In this work, we have explored the possibility of preserving or
enhancing entanglement between two atoms after their interaction
with a thermal field. We have investigated the interaction for two
kinds of initial atomic states
$\ket{\Phi_\beta}=1/\sqrt{2}(\sin\beta\ket{01}+\cos\beta\ket{10})$
and
$\ket{\Psi_\beta}=1/\sqrt{2}(\sin\beta\ket{00}+\cos\beta\ket{11})$.
Although this physical system had been discussed previously, new
interesting phenomena have been found and the already reported
results have been confirmed. For the initial state
$\ket{\Phi_\beta}$, we observe a very different behavior after the
interaction with the thermal field for two distinguished intervals
of input states. Qualitatively, for almost the whole range
$\beta=(0,\pi/2)$, the EOF is decreased while for the remaining
input states there are oscillations in the EOF so that they are
never smaller than their initial value. This result is more
significant at low temperatures, hence, the entanglement
enhancement mechanism is induced by spontaneous decay.

For the second family of input states parameterized by
$\ket{\Psi_\beta}$, there is no entanglement enhancement for
almost any $\beta$. In this case, the entanglement is generally
degraded matching our intuitive conception of an interaction
between a correlated pairs of qubits and thermal light.
Nevertheless, the output atomic density matrix can be decomposed
into $\ket{\phi}=1/\sqrt{2}\ket{01+10}$ and a partially entangled
mixed state, both of them living in mutually orthogonal subspaces.
Thus they can be discriminated with zero probability of error. If
we consider that this holds for all $\beta$ when $\bar n$ is
sufficiently high, it follows that we are able to extract the
maximally entangled state $\ket{\phi}$ even if the input is a
factorized state and the output is in a separable form. We may
conclude that we have at disposal a probabilistic source of
maximally entangled atomic pairs, when our deposit is just a
sufficiently hot thermal light with two atoms in a factorized
state.

In current experiments with trapped ions, the energies involved in
the transitions compared with room temperatures are such that
$\hbar\Omega/kT\ll1$ for the Innsbruck group ($729$ nm atomic
transition in $^{40}$Ca$^+$)~\cite{ion2_nature} and
$\hbar\Omega/kT\gg1$ for the NIST group ($^9$Be$^+$ at $1.2$
GHz)~\cite{ion1_nature}. Thus, one would expect that the former is
a good scenario for testing the predictions made for the initial
state $\ket{\Phi_\beta}$, in particular, the entanglement
enhancement. Meanwhile, the NIST group parameters could be
appropriate to study the possibility of stabilized postselection
of the maximally entangled state described for the interaction of
thermal resonant light with $\ket{\Psi_\beta}$.

\begin{acknowledgments}
The authors are grateful to S.~Hacyan for valuable comments.
Funding from CONACyT $41048$-F is acknowledged.
\end{acknowledgments}

\appendix*
\section{}
\subsection*{Initial Atomic State $\ket{\Phi_\beta}$}

With the help of the inverse transformation from the bare state to
the dressed state basis, we rewrite the initial density matrix
determined by the atomic state $\ket{\Phi_\beta}$ and a thermal
single-mode field
\begin{eqnarray}\label{ro_zero}
    \varrho^{(\Phi)}(\beta,t=0) &=&
    \sum_{n=0}^\infty{\frac{p_n}{4}}
    \left[
    \left(1+\sin2\beta\right)(-\ket{3}_n+\ket{4}_n)(-\bra{3}_n+\bra{4}_n)\right.\nonumber\\
    &&\left.-\sqrt{2}\cos
    2\beta\left((-\ket{3}_n+\ket{4}_n)\bra{2}_n+\ket{2}_n(-\bra{3}_n+\bra{4}_n)\right)
    -2 (1-\sin 2\beta)\ketbra{2}{2}_n\right],
\end{eqnarray}
where $p_n=[1-\exp{(-\hbar\Omega/kT)}]\exp{(-n\hbar\Omega/kT)}$.
The time evolution of the whole system is directly found. Tracing
over the resulting density matrix we obtain an output atomic state
(Eqs.~(\ref{rho_Phi})) with elements
\begin{align}\label{ro_param_Phi}
   a_1 &=
   \frac{1}{8}(1+\sin2\beta)\left(m_+-h_2(2t)-\frac{1}{2}h_1(2t)
   \right)\\
   a_2 &=
    \frac{1}{8}(1+\sin2\beta)\left(1+h_2(2t)\right)
   -\frac{1}{2}h_2(t)\cos 2\beta
   +\frac{1}{4}(1-\sin2\beta)\\
   a_3 &=
   \frac{1}{8}(1+\sin2\beta)\left(1+h_2(2t)\right)
   -\frac{1}{4}(1-\sin2\beta)\\
   a_4 &=
   \frac{1}{8}(1+\sin2\beta)\left(1+ h_2(2t)\right)
   +\frac{1}{2}h_2(t)\cos 2\beta
   +\frac{1}{4}(1-\sin2\beta)\\
   a_5 &=
   \frac{1}{8}(1+\sin2\beta)\left(m_--h_2(2t)+\frac{1}{2}h_1(2t)
   \right)
\end{align}
with
\begin{subequations}\label{hfunctions}
\begin{eqnarray}
h_1(t)
&=&\sum_{n=0}^\infty{\frac{p_n}{n+1/2}}\cos\alpha_{n-1}t\\
h_2(t) &=&\sum_{n=0}^\infty p_n\cos\alpha_{n-1}t =
-\frac{1}{4g^2}h_1^{\prime\prime}(t).
\end{eqnarray}
\end{subequations} $\alpha_{n-1}=2g\sqrt{n+1/2}$ is the Rabi
frequency coming from eigenvalues of the interaction
Hamiltonian~(\ref{Hint}) and $m_\pm=1\pm
M\arctan\left(\exp\left(-{\hbar\Omega/2kT}\right)\right)$ where
$M=\left[1-\exp\left(-{\hbar\Omega/kT}\right)\right]
\exp\left({\hbar\Omega/ 2kT}\right)$.

Expressions~(\ref{hfunctions}) appearing in the matrix elements
can be written in a more compact form
\begin{equation}
h_2(t)=2\sinh\kappa \sum_{n=0}^\infty
e^{\kappa(2n+1)}\cos\sqrt{2n+1}gt =2\sinh2\kappa\sum_{n=0}^\infty
e^{\kappa n}\cos\sqrt{n}gt -\sum_{n=0}^\infty e^{2\kappa
n}\cos\sqrt{2n}gt
\end{equation}
where $\kappa=\frac{\hbar\Omega}{2kT}$. Using the Abel-Plana
formula
\begin{equation}
\sum_{n=0}^\infty F(n) =\int_0^\infty F(x){\rm d}x+\frac{1}{2}F(0)
+i\int_0^\infty\frac{F(ix)-F(-ix)}{e^{2\pi x}-1}{\rm d}x
\end{equation}
the following  integral expression for $h_2(t)$ is found
\begin{equation}
h_2(t)=2\frac{\sinh\kappa}{\kappa}\left[\int_0^\infty
e^{-x}\cos\tilde t\sqrt{x}{\rm d}x +4\int_0^\infty\frac{\sin
x\cosh \tilde t\sqrt{x} -\sin 2x\cosh \tilde
t\sqrt{2x}}{e^{\frac{\pi x}{\kappa}}-1}{\rm d}x\right]
\end{equation}
with an adimensional variable $\tilde t=\frac{gt}{\sqrt{\kappa}}$
that establishes a natural time scale
$\tau_0=\frac{\sqrt{\kappa}}{g}$ for the description of the
interaction. Notice that the second integral is an exponential
decreasing function of the temperature so that for a sufficiently
hot environment
\begin{equation}
h_2(t)\sim2\frac{\sinh\kappa}{\kappa}\int_0^\infty
e^{-x}\cos\tilde t\sqrt{x}{\rm d}x
=2\frac{\sinh\kappa}{\kappa}\left[1-\tilde t e^{-\tilde
t^2}\int_0^{\tilde t}e^{x^2}{\rm d}x\right] .
\end{equation}
The latter integral can be recognized as an error function with an
imaginary argument. An integral expression for $h_1(t)$ can be
found in a similar manner.

\subsection*{Initial Atomic State $\ket{\Psi_\beta}$}

Calculations similar to the case of the initial state
$\ket{\Phi_\beta}$ lead to the following matrix coefficients of
Eq.~(\ref{rho_Psi})
\begin{align}\label{ro_param_Psi}
    b_1 &=\sum_{n=0}^\infty p_n\left[
      \sin^2\beta
      \left(\frac{n-1}{2n-1} +\frac{n}{2n-1}\cos\alpha_{n-2}t\right)^2
     +\cos^2\beta\frac{(n+1)(n+2)}{(2n+3)^2}(1-\cos\alpha_n t)^2
     \right]
  \\
    b_2 &=e^{2i\Omega t}\sin\beta\cos\beta\sum_{n=0}^\infty p_n
      \left(\frac{n-1}{2n-1}
      +\frac{n}{2n-1}\cos\alpha_{n-2}t\right)\left(\frac{n+2}{2n+3}
      +\frac{n+1}{2n+3}\cos\alpha_nt\right)
  \\
    b_3 &=\sum_{n=0}^\infty p_n\left[
      \sin^2\beta\frac{n}{4n-2}\sin^2\alpha_{n-2}t
      +\cos^2\beta\frac{n+1}{4n+6}\sin^2\alpha_{n}t
      \right]
  \\
    b_4 &=\sum_{n=0}^\infty p_n\left[
      \sin^2\beta
      \frac{n(n-1)}{(2n-1)^2}(1-\cos\alpha_{n-2}t)^2+\cos^2\beta
      \left(\frac{n+2}{2n+3} + \frac{n+1}{2n+3}\cos\alpha_nt\right)^2
      \right].
\end{align}


\begin{thebibliography}{99}

\bibitem{ion1_nature}Leibried D,  Knill E,  Seidelin S,  Britton J,
Blakestad R B, Chiaverini J,  Hume D B, Itano W M, Jost J D,
Langer C,  Ozeri R,  Reichle R, and  Wineland D J 2005 {\it
Nature} {\bf438}, 639

\bibitem{ion2_nature}H\"afner H, H\"ansel W, Roos C F, Benhelm J,
Chek-al-kar D, Chwalla M,  K\"orber T, Rapol U D, Riebe M, Schmidt
P O , Becher C, G\"uhne O, D\"ur W, and Blatt R 2005 {\it Nature}
{\bf438}, 643

\bibitem{thermal_light}Kim M S, Lee J, Ahn D, and Knight P L 2002 {\it Phys. Rev.} A {\bf65},
040101

\bibitem{thermal_light_ent}Bose S, Fuentes-Guridi I, Knight P L, and
Vedral V 2001 {\it Phys. Rev. Lett.} {\bf87}, 050401

\bibitem{thermal_light_ent1}An J-H, Wang S-J, and Luo H-G 2005 {\it J. Phys.} A {\bf38}, 3579


\bibitem{fock}Sainz I, Klimov A B, and Roa L 2006 {\it Phys. Rev. }A {\bf73},
032303; Braun D 2002 {\it Phys. Rev. Lett.} {\bf89}, 277901;
Kudryavtsev I K, Lambrecht A, Moya-Cessa H and Knight P L 1993
{\it J. Mod. Optics} 40, 1605; Tessier T E, Deutsch I H, Delgado A
and Fuentes-Guridi I 2003 {\it Phys. Rev. }A {\bf68}, 062316

\bibitem{whitenoise}Plenio M B and Huelga S F  2002 {\it Phys. Rev. Lett.} {\bf88}, 197901


\bibitem{thermal_others}Schneider S and Milburn G J 2002 {\it Phys. Rev. }A {\bf65}, 042107;
Lucamarini M, Paganelli S, and Mancini S 2004 {\it Phys. Rev. }A
{\bf69}, 062308

\bibitem{tavis}Tavis M and Cummings F W 1967 {\it Phys. Rev. }{\bf170},
379

\bibitem{EOF}Bennett C H, DiVincenzo D P, Smolin J A, and
Wootters W K 1996 {\it Phys. Rev. }A {\bf54}, 3824

\bibitem{wootters}Hill S and Wootters W K 1997 {\it Phys. Rev. Lett.} {\bf78}, 5022;
Wootters W K 1998 {\it Phys. Rev. Lett.} {\bf80}, 2245


\bibitem{measures}Horodecki M, Horodecki P, and Horodecki R 2000
{\it Phys. Rev. Lett.} {\bf84}, 2014; Vedral V, Plenio M B, Rippin
M A , and Knight P L 1997 {\it Phys. Rev. Lett.} {\bf78}, 2275

\bibitem{plenio}Plenio M B and  Virmani S, quant-ph/0504163

\bibitem{negativity} \.Zyczkowski K,  Horodecki P,  Sanpera A, and
Lewenstein M 1998 {\it Phys. Rev. }A {\bf58}, 883;  Vidal G and
Werner R F 2002 {\it Phys. Rev.} A {\bf65}, 032314

\bibitem{uniqueness}Popescu S and Rohrlich D 1997 {\it Phys. Rev. }A {\bf56},
R3319; Donald M J,  Horodecki M, and  Rudolph O 2002 {\it J. Math.
Phys.} {\bf43}, 4252

\bibitem{CREN} Lee S, Chi D P, Oh S D, and Kim J 2003 {\it Phys. Rev.} A {\bf68}, 062304

\bibitem{ent_exp_measure1} Ac\'in A,  Tarrach R, and  Vidal G 2000
{\it Phys. Rev.} A {\bf61}, 062307

\bibitem{ent_exp_measure2}Horodecki P 2003 {\it Phys. Rev. Lett.} {\bf90}, 167901


\end{thebibliography}
\end{document}